\documentclass[twocolumn]{aastex6}
\usepackage{amsmath,amstext}
\usepackage[T1]{fontenc}
\usepackage{apjfonts}
\usepackage[figure,figure*]{hypcap}
\usepackage{hyperref}

\begin{document}

\title{On the minimum jet power of TeV BL Lac objects in the p-$\gamma$ model}

\author{Rui Xue\altaffilmark{1}, Ruo-Yu Liu\altaffilmark{2}, Xiang-Yu Wang\altaffilmark{1}, Huirong Yan\altaffilmark{2,3} and Markus B\"ottcher\altaffilmark{4}}
\altaffiltext{1}{School of Astronomy and Space Science, Nanjing University, Nanjing 210093, China}
\altaffiltext{2}{Deutsches Elektronen Synchrotron (DESY), Platanenallee 6, D-15738 Zeuthen, Germany}
\altaffiltext{3}{Institut f\"ur Physik und Astronomie, Universit\"at Potsdam, D-14476 Potsdam, Germany}
\altaffiltext{4}{Centre for Space Research, North-West University, Potchefstroom 2520, South Africa}


\begin{abstract}
We study the requirement on the jet power in the conventional $p-\gamma$ models (photopion production and Bethe-Heitler pair production) for TeV BL Lac objects. 
We select a sample of TeV BL Lac objects whose SEDs are difficult to be explained by the one-zone leptonic model. Based on the relation between the 
$p-\gamma$ interaction efficiency and the opacity of $\gamma\gamma$ absorption, we find that the detection of TeV emission poses upper limits on the $p-\gamma$ interaction efficiencies in these sources and hence minimum jet powers can be derived accordingly. We find that the obtained minimum jet powers exceed the Eddington luminosity of the supermassive black holes. Implications for the accretion mode of the supermassive black hole in these BL Lac objects and the origin of their TeV emissions are discussed.
\end{abstract}


\maketitle

\section{Introduction}

Blazars are the most extreme form of active galactic nuclei (AGN), with their jets pointing in the direction of the observer \citep{1995PASP..107..803U}. 
Multi-wavelength observations show that the spectral energy distributions (SEDs) of blazars generally exhibit a two-bump structure. The origin of the 
low-energy bump is generally considered to be synchrotron radiation of relativistic electrons accelerated in the jet, while the origin of the high-energy 
bump is still under debate. In leptonic models, the high-energy bump is explained as inverse Compton (IC) scattering, in which the high-energy electrons 
in the jet up-scatter the low energy photons from the external photon field such as the emission of the broad line region or the accretion disk (external 
Compton, EC), or the synchrotron radiation of the electrons of the same population (i.e., synchrotron-self Compton, SSC). In hadronic models, the 
high-energy bump is instead assumed to originate from proton-synchrotron emission, or emission from secondary particles generated in photohadronic 
and Bethe-Heitler (BH) interactions \citep{1993A&A...269...67M, 2000NewA....5..377A, 2003ApJ...586...79A}. Hereafter, we denote the photohadronic and 
BH interactions collectively by $p-\gamma$ interactions.

So far, 71 blazars have been detected in the TeV band, most of which are high-synchrotron-peaked BL Lac objects (HBLs)\footnote{http://tevcat.uchicago.edu/}. Due to the lack of the strong emission from the external photon field in BL Lac objects, the SSC model is usually employed to explain the high-energy emissions in the leptonic model \citep{1997A&A...320...19M, 2000ApJ...536..742P, 2002MNRAS.336..721K, 2014MNRAS.439.2933Y}. The TeV emission from blazars is absorbed due to the $\gamma\gamma$ pair production by interacting with the extragalactic background light (EBL). After correcting for EBL absorption, the intrinsic TeV 
spectrum is harder than the observed one. Particularly, the TeV spectra of some HBLs are too hard to be explained with the SSC mechanism, since the 
Klein-Nishina (KN) effect softens the IC spectrum in the TeV band. Thus, the hard TeV spectra pose a challenge to the leptonic explanation and may 
suggest a hadronic origin.

Among hadronic models, the proton synchrotron model is often employed. In general, proton synchrotron spectra typically peak at multi-GeV rather than 
TeV energies \citep{2013ApJ...768...54B, 2018arXiv180702085P}, unless one  considers extreme scenarios, such as protons with energy $10^{20}\,$eV radiating 
in kilo-Gauss magnetic fields, or Doppler factors of $\sim 100$ for the jet. 
Also, some studies \citep{2015MNRAS.450L..21Z, 2016ApJ...825L..11P} suggest that the minimum jet power in the proton synchrotron model will 
exceed the Eddington luminosity of the supermassive black hole (SMBH) which launches the jet. In the $p-\gamma$ model, the energy of relativistic 
protons is mainly lost through photohadronic interactions ($p + \gamma \rightarrow p/n + \pi^0 + \pi^\pm$) and BH pair 
production ($p + \gamma \rightarrow p + e^{\pm}$), with the radiation zone being relatively compact. However, \cite{2009ApJ...704...38S} 
and \cite{2011IAUS..275...59S} argue that the $p-\gamma$ interactions are very inefficient in flat spectrum radio quasars (FSRQs) so that an 
extremely high proton power is required in order to explain the high energy radiation with $p-\gamma$ interactions.

It should be noted that there is a robust connection between the efficiency of $p-\gamma$ interactions and the opacity of the internal 
$\gamma\gamma$ pair production in $p-\gamma$ models, since  the target photon fields of these processes are the same. It has been 
shown that the interaction efficiency of photohadronic processes in the high-energy limit is about 1000 times smaller than the peak 
$\gamma\gamma$ opacity \citep{2000NewA....5..377A, 2007ApJ...664L..67D, 2012ApJ...755..147D}. Such a relation implies that if $p-\gamma$ interactions are 
very efficient, high-energy gamma-ray emission should not be expected to be detected from the same object. On the other hand, the detection of 
high-energy gamma-ray emission from certain BL Lac objects can in turn place an upper limit on the efficiency of $p-\gamma$ interactions, 
which then translates to a minimum proton power in the jet. In this work, we will derive conservative yet robust lower limits on jet 
powers based on observations of some TeV BL Lac objects, utilizing the relation between the $p-\gamma$ interaction efficiency and the 
internal $\gamma\gamma$ pair production opacity. Note that any emission from $p-\gamma$ interactions, despite the complicated 
electromagnetic cascade induced by secondary particles, eventually originates from the energy of protons lost in $p-\gamma$ 
interactions. Thus, the constraint arising from the opacity of the $\gamma\gamma$ annihilation applies to any model in the 
framework of $p-\gamma$ processes, no matter which radiation mechanism (e.g., synchrotron, IC, pionic radiation) or which type of 
radiating particles (e.g., electron/positron, muon, neutral pion) are involved.

The rest of this paper is structured as follows. In Section~\ref{model} we describe our method to obtain a lower limit on the proton power of an AGN jet. 
We apply our method to a sample of 9 TeV BL Lacs in Section~\ref{app}; in Section~\ref{dis} we present our discussion and conclusions. Throughout the paper, the $\Lambda$CDM cosmology with $H_{\rm{0}}={70\rm{km~s^{-1} Mpc^{-1}}}$, $\Omega_{\rm{m}}=0.3$, $\Omega_{\rm{\Lambda}}=0.7$ is adopted.\label{intro}

\section{Model description}\label{model}
\subsection{The injected particle energy distribution}
In BL Lac objects, the target photon field for the $p-\gamma$ interactions and the $\gamma\gamma$ pair production is mainly provided by the synchrotron 
radiation of electrons. The proton spectral shape is crucial to the overall radiation efficiency of protons. Thus, we firstly model the electron and 
proton spectrum in the radiation zone of the jet. The parameters are measured in the comoving frame of the jet unless otherwise specified.

We assume a single spherical radiation zone of radius $R$ being composed of a plasma of electrons and protons in a uniformly entangled magnetic field 
($B$), and the observed emission is boosted by a relativistic Doppler factor $\delta_{\rm D}$. Assuming the jet moves with a bulk Lorentz factor 
$\Gamma$ (or with a velocity of $\beta=\sqrt{1-1/\Gamma^2}$ in units of the speed of light $c$), we have $\delta_{\rm D}=[\Gamma(1-\beta \rm cos\theta)]^{-1}\approx \Gamma$ for a relativistic jet close to the line of sight in blazars with a viewing angle of $\theta \lesssim 1/\Gamma$. To explain the low-energy bump in the SED, a broken power-law distribution
is required for the electron injection spectrum, i.e.
\begin{equation}
Q_{e}(\gamma_{e}) = Q_{\rm e, 0}\gamma_{\rm e}^{-q_{1}}\left[1 + \left(\frac{\gamma_{e}}{\gamma_{e,\rm b}}\right)^{(q_2 - q_1)}\right]^{-1}, 
\gamma_{e,\rm  min} < \gamma_{e} < \gamma_{e,\rm max},
\end{equation}
where $Q_{e, 0}$ is the normalization, $\gamma_{e}$ is the electron Lorentz factor, $\gamma_{e,\rm min}$ is the minimum Lorentz factor, 
$\gamma_{e,\rm max}$ is the maximum electron Lorentz factor, $\gamma_{e, \rm b}$ is the break electron Lorentz factor, $q_1$ and $q_2$ 
are the spectral indices below and above $\gamma_{e, \rm b}$. Given an electron injection luminosity $L_{e, \rm inj}$ in the blob, $Q_{e,0}$ 
can be obtained by $\int Q_{e}\gamma_{e} m_{e} c^2d\gamma_{e}=L_{e, \rm inj}/(4/3\pi R^3)$ where $m_e$ is the mass of an electron. We assume 
a quasi-steady state is reached, and the injection is balanced by radiative cooling and/or particle escape. The number density 
of the injected electrons in the radiation zone can be obtained by $Q_{e}t_{e}$, where $t_{e} = min\{ t_{\rm cool}, t_{\rm dyn}\}$. 
$t_{\rm cool} = \frac{3m_{e}c}{4(U_{\rm B}+\kappa_{\rm KN}U_{\rm ph})\sigma_{\rm T}\gamma_{e}}$ is the radiative cooling time where 
$U_{\rm B} = \frac{B^2}{8\pi}$ is the energy density of the comoving magnetic field, $U_{\rm ph}$ is the energy density of the soft 
photons, $\sigma_{\rm T}$ is the Thomson scattering cross section and $\kappa_{\rm KN}$ is a numerical factor accounting for the KN 
effect. $t_{\rm dyn}$ is the dynamical timescale of the blob, which may be determined by the adiabatic expansion of the blob or by 
the particle injection processes. Typically, we have $t_{\rm dyn}\simeq R/c$. Then, the quasi-steady electron spectrum in the blob 
shows a double broken power-law form \citep{1996ApJ...463..555I}, i.e.
\begin{equation}
\begin{split}
N_{\rm e}(\gamma_{\rm e}) = N_{\rm e, 0}\gamma_{\rm e}^{-q_{1}}\left[1 + \left(\frac{\gamma_{\rm e}}{\gamma_{\rm e, b}}\right)^{(q_2 - q_1)}\right]^{-1}\\
\times(1 + \frac{\gamma_{\rm e}}{\gamma_{\rm cool}})^{-1}, \gamma_{\rm e, min} < \gamma_{\rm e} < \gamma_{\rm e, max},
\end{split}
\end{equation}
where $N_{\rm e, 0}$ is the normalization coefficient which is equal to $Q_{\rm e,0}t_{\rm ad}$, and 
$\gamma_{\rm cool} = \frac{3m_{e}c^2}{4(U_{\rm B} + U_{\rm ph})\sigma_{\rm T}R}$ is the electron Lorentz factor where $t_{\rm ad} = t_{\rm cool}$. 
Then the kinetic power in relativistic electrons $P_{\rm e, k}$ in the AGN frame is given by
\begin{equation}
P_{\rm e, k} \simeq \pi R^2\Gamma^2 m_{\rm e}c^3\int \gamma_{\rm e}N_{\rm e}(\gamma_{\rm e})d\gamma_{\rm e}.
\end{equation}

Protons are assumed to be injected with a power-law distribution\footnote{Note that if we assume the proton spectral shape is same as the electron spectral shape, which was a broken power law, the overall $p-\gamma$ interaction efficiency will be decreased, since the amount of protons at high energy, where the $p-\gamma$ interaction efficiency is relatively high, will be reduced significantly. As a result, the required energy budget for relativistic protons will be even higher. Please refer to the discussion in later sections.}, i.e.
\begin{equation}
Q_{\rm p}(\gamma_{\rm p}) = Q_{\rm p, 0}\gamma_{\rm p}^{-q}, \gamma_{\rm p, min} < \gamma_{\rm p} < \gamma_{\rm p, max},
\end{equation}
where $Q_{\rm p, 0}$ is the proton injection constant, $\gamma_{\rm p}$ is the proton Lorentz factor, $q$ is the spectral index, $\gamma_{\rm p, min}$ 
is the minimum proton Lorentz factor which is usually $\sim 1$ and $\gamma_{\rm p, max}$ is the maximum proton Lorentz factor which can be obtained by 
comparing the acceleration timescale and the escape timescale of protons. {For a proton injection rate $Q_{\rm p}(\gamma_{\rm p})$ per unit volume 
in the blob, the quasi-steady-state proton energy distribution is given by
\begin{equation}
N_{\rm p}(\gamma_{\rm p}) \approx t_{\rm dyn} \, Q_{\rm p}(\gamma_{\rm p}),
\end{equation}
since protons generally are not cooled efficiently in the $p-\gamma$ model. Therefore, we can obtain the kinetic power in relativistic protons as}
\begin{equation}\label{eq:Ppk}
P_{p,k}\simeq \pi R^2\Gamma^2 m_{\rm p}c^3\int \gamma_{\rm p}N_{\rm p}(\gamma_{\rm p})d\gamma_{\rm p}= \pi R^3\Gamma^2 m_{\rm p}c^2\int 
\gamma_{\rm p}Q_{\rm p}(\gamma_{\rm p})d\gamma_{\rm p}
\end{equation}
We assume that particle acceleration is dominated by diffusive shock acceleration, for which the acceleration timescale in the relativistic limit 
can be evaluated by 
\citep{2004PASA...21....1P, 2007Ap&SS.309..119R}
\begin{equation}
t_{\rm acc}\simeq \frac{3\alpha}{20} \frac{r_{\rm L}}{c} \simeq \frac{20\alpha}{3} \frac{\gamma_{\rm p}m_{\rm p}c}{eB}
\end{equation}
under the quasi-linear theory, where $r_{\rm L}$ is the Larmor radius of the proton
and $\alpha$ is the ratio of the mean magnetic field energy density to the turbulent magnetic field energy density. 
Generally, we expect $\alpha=10-100$ or even larger \citep[][but also see e.g. \citealt{Bell04} for the discussion on saturation of turbulent magnetic field]{1983A&A...125..249L, 1984ARA&A..22..425H}, but in the very limiting case the value of $\alpha$ may approach unity. On the other hand, we 
assume protons escape via diffusion so the escape timescale can be written as
\begin{equation}
t_{\rm esc} = \frac{R^2}{4D}=\frac{3eBR^2}{4\alpha \gamma_{\rm p} m_{\rm p} c^3},
\end{equation}
where $D$ is the diffusion coefficient, $D=\alpha r_{\rm L}c/3$. Of course, there may be other processes which could play a more 
important role in limiting the acceleration, such as the advective escape or adiabatic cooling. A more sophisticated treatment requires detailed modeling 
of,e.g., the geometry and the configuration of the magnetic field. We here simply consider diffusion as the main escape mechanism so that we can obtain 
the maximum proton Lorentz factor as
\begin{equation}
\gamma_{p,\rm max}=\sqrt{\frac{80}{9}}\frac{eBR}{\alpha m_pc^2}\simeq 10^8\left(\frac{\alpha}{10}\right)^{-1}
\left(\frac{B}{1\rm G}\right)\left(\frac{R}{10^{16}\rm cm}\right)
\end{equation}

\subsection{The minimum injection proton power}
Due to the large number of free parameters in $p-\gamma$ models, it is generally impossible to obtain a unique set of parameters by modeling the SED. 
However, it is meaningful to search for the minimum jet power and compare it with the Eddington luminosity of the SMBH. In this subsection, we 
propose a method of obtaining a robust lower limit on the proton power in the $p-\gamma$ model. 

\begin{figure*}[htbp]
\centering
\includegraphics[angle=0.2, scale=0.31]{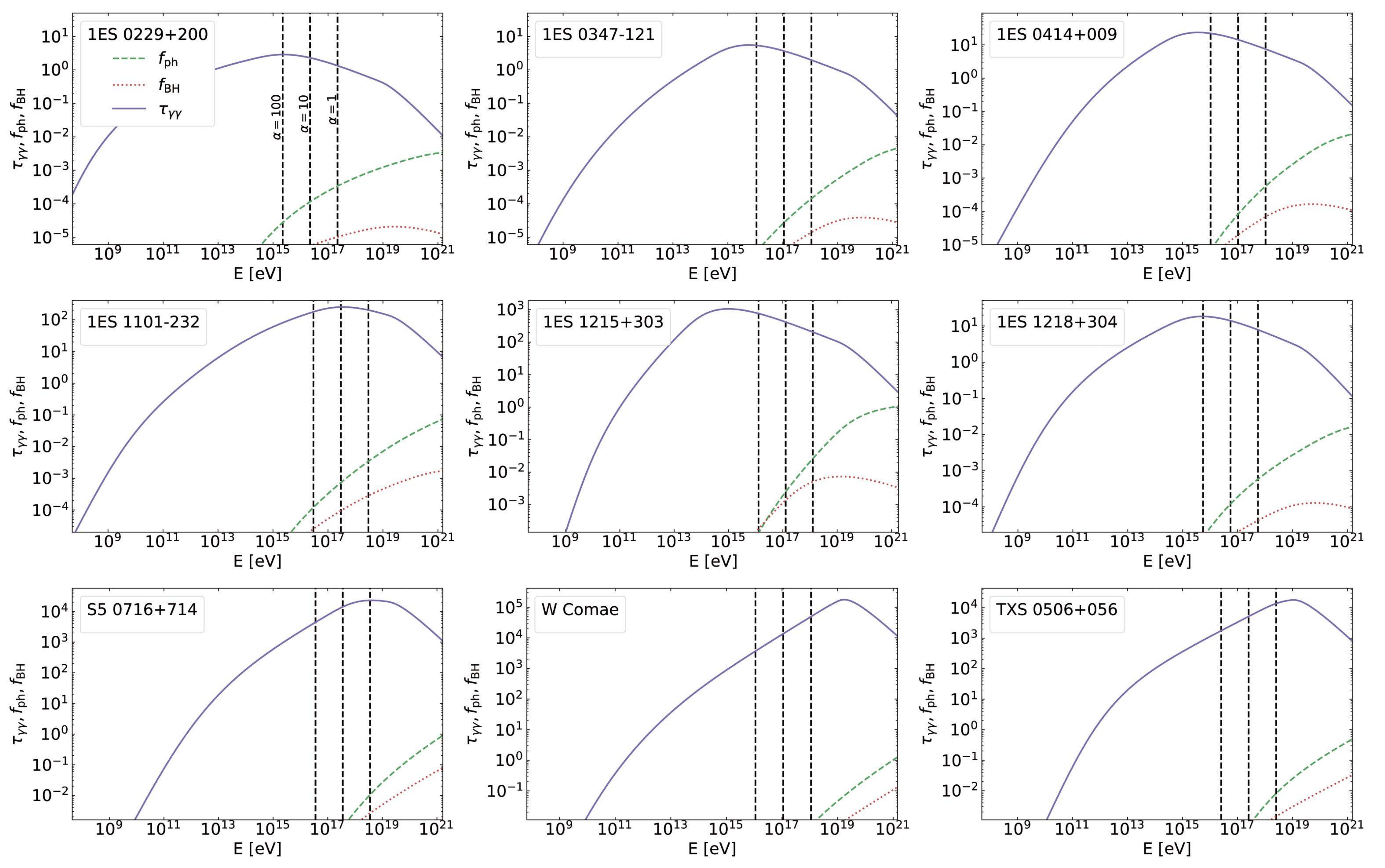}\label{fig:efficiency}
\caption{Correlation between $f_{\rm ph}$, $f_{\rm BH}$ and $\tau_{\gamma\gamma}$ when $\tau_{\rm \gamma\gamma}(E_{\rm c})=1$. 
The purple solid curve shows $\tau_{\gamma\gamma}$ as a function of the high energy photon energy in the comving frame. The green dashed and red dotted curves show $f_{\rm ph}$ and $f_{\rm BH}$ as functions of the relativistic proton energy in the comving frame. In each panel, three vertical dashed lines represent $\gamma_{\rm p, max}m_{\rm p}c^2$ when $\alpha=1, 10$ and 100, respectively, from left to right.}
\end{figure*}

The efficiency of photohadronic interactions and BH (Bethe-Heitler) pair production in a radiation field can be written as \citep{1968PhRvL..21.1016S, 1990acr..book.....B}
\begin{equation}\label{eq:eff_ph}
f_{\rm ph}(\gamma_{\rm p}) = \frac{t_{\rm dyn}}{t_{\rm ph}} =\frac{R}{2\gamma_{\rm p}^2} \int_{\epsilon_{\rm th}^{\rm ph}/{2\gamma_{\rm p}}}^{\infty} 
d\epsilon \frac{n_{\rm ph}(\epsilon)}{\epsilon^2} \int_{\epsilon_{\rm th}^{\rm ph}}^{2\epsilon\gamma_{p}} d\epsilon_{r} 
\sigma_{\rm ph}(\epsilon_{r})K_{\rm ph}(\epsilon_{r})\epsilon_{r}
\end{equation}
and
\begin{equation}\label{eq:eff_bh}
f_{\rm BH}(\gamma_{\rm p}) = \frac{t_{\rm dyn}}{t_{\rm BH}} = \frac{R}{2\gamma_{\rm p}^2} \int_{\epsilon_{\rm th}^{\rm BH}/{2\gamma_{\rm p}}}^{\infty} 
d\epsilon \frac{n_{\rm ph}(\epsilon)}{\epsilon^2} \int_{\epsilon_{\rm th}^{\rm BH}}^{2\epsilon\gamma_{p}} d\epsilon_{r} \sigma_{\rm BH}(\epsilon_{r})
K_{\rm BH}(\epsilon_{r})\epsilon_{r},
\end{equation}
respectively, where $\epsilon$ represents the photon energy in the jet frame, and $\epsilon_{\rm th}^{\rm ph / \rm BH}$ is the photon threshold energy in the rest frame of proton for the photohadronic and BH processes, respectively, $n_{\rm ph}(\epsilon)$ is the number density of the soft photons in the comving frame which is mainly provided by the synchrotron radiation of primary electrons, $\epsilon_{r}$ is the photon energy in the rest frame of proton, $\sigma_{\rm ph}(\epsilon_{r})$ \citep{2000CoPhC.124..290M} is the cross section for photopion production, $\sigma_{\rm BH}(\epsilon_{r})$ \citep{1992MNRAS.259..218C} is the cross section for BH pair production, $K_{\rm ph}(\epsilon_{\rm r})$ \citep{2000CoPhC.124..290M} is the inelasticity of photohadronic interactions and $K_{\rm BH}(\epsilon_{\rm r})$ \citep{1992MNRAS.259..218C} is the inelasticity of BH pair production.

Generally, the high energy photons produced in the jet will be attenuated by interacting with the synchrotron radiation of primary electrons in the jet. 
This internal $\gamma\gamma$ absorption optical depth can be calculated as \citep{2008ApJ...686..181F, 2009herb.book.....D}
\begin{equation}
\tau_{\rm \gamma\gamma}(\epsilon_1) =\frac{t_{\rm dyn}}{t_{\gamma\gamma}} =\frac{2R(m_ec^2)^4}{\epsilon_1^2}\int_{m_e^2c^4/\epsilon_1}^{\infty} d\epsilon 
\frac{n_{\rm ph}(\epsilon)}{\epsilon^2}\int_1^{\frac{\epsilon\epsilon_1}{m_e^2c^4}}ds\sigma_{\rm \gamma\gamma}(s)s,
\end{equation}
where $\sigma_{\rm \gamma\gamma}$ is the $\gamma\gamma$ pair-production cross section, $\sqrt{s}$ is the center-of-momentum frame Lorentz factor of the 
produced electron and positron \citep{2009herb.book.....D}, and $\epsilon_1$ is the energy of $\gamma$-ray photons.

To get efficient hadronic emission, we may in principle adjust model parameters to result in a large $f_{p\gamma}(\equiv f_{\rm ph}+f_{\rm BH})$. However, 
$\tau_{\rm \gamma\gamma}$ will be increased simultaneously since the target photon fields for $\gamma\gamma$ absorption and for $p-\gamma$ processes are 
the same. Here we define a critical energy $E_{\rm c}$ in the AGN frame beyond which the TeV spectrum of a BL Lac shows a cutoff or softening. If there 
is no such feature in the TeV spectrum then $E_{\rm c}$ is defined to be equal to the highest energy that the TeV detection extends to. Assuming 
$\tau_{\rm \gamma\gamma}(E_{\rm c})=1$ for the BL Lac object, we can obtain an upper limit on $f_{p\gamma}$. A simplified expression for this relation 
can be obtained using the $\delta$-approximation for the cross sections of both processes. In this approximation, the energies of the soft photon $\epsilon_s$ 
and the high energy photon $E_{\rm c}$ in the observer's frame satisfy the relation $\epsilon_s E_{\rm c}\approx 4\delta_{\rm D}^2m_e^2c^4$ for the $\gamma\gamma$ 
annihilation, while the same photons interact  with protons of energy $E_{p,c}\simeq \delta_D^2m_pc^2\epsilon_\Delta/\epsilon_s$ where 
$\epsilon_\Delta\simeq 0.3\,$GeV is the energy for the $\Delta$ resonance. Thus, we have $E_{p,c}\simeq 3\times 10^5E_{\rm c}$ which does not depend on any 
model parameters. If we compare the peak cross section for $\gamma\gamma$ annihilation $\sigma_{\rm \gamma\gamma,\rm peak}\simeq 10^{-25}\rm\,cm^{-2}$ 
with the peak value of the product of the cross section and the inelasticity for the photopion production (i.e.,$\sim 10^{-28}\rm \,cm^{-2}$), we obtain 
the relation. 
\begin{equation}
f_{\rm ph}(E_{p,c})\simeq 10^{-3}\tau_{\gamma\gamma}(E_{\rm c}).
\end{equation}
Thus, the condition $\tau_{\rm \gamma\gamma}(E_{\rm c})=1$ suppresses the $p-\gamma$ interaction efficiency to a quite low level, i.e., 
$f_{\rm ph}(E_{p, c}) \sim 10^{-3}$. For protons with other energies, $f_{p\gamma} = f_{p\gamma}(E_{p,c}) F(E_p)$, where $F (E_p)$ is a normalized function 
depicting how $f_{p\gamma}$ changes with $E_p$. Since we fix $\tau_{\gamma\gamma}(E_{\rm c}) = 1$ in our treatment, the uncertainty of $f_{p\gamma}$ mainly originates from the uncertainty of $F(E_p)$ which is determined by the spectral shape of the target photon field for $p-\gamma$ interactions, i.e., the spectral shape of the electron synchrotron radiation. Although the low-energy SED may be fitted with different combinations of parameters, the resulting spectral shape should be always compatible with the observation from the optical band to the X-ray band. Thus, $F(E_p)$ is more or less fixed for a given source. In addition, the condition $\tau_{\gamma\gamma}(E_{\rm c}) = 1$ also reduces the degeneracy of the model parameters. Therefore, $F(E_p)$ is not expected to vary significantly. A similar relation can be also obtained for the BH process. Fig.~\ref{fig:efficiency} shows the the $f_{\rm ph}$, $f_{\rm BH}$ and $\tau_{\gamma\gamma}$ as functions of proton/photon energy. 

Here we take 1ES 0229+200 as an example to further interpret the relation between $\tau_{\rm \gamma\gamma}$ and $f_{\rm ph}$. The critical energy $E_{\rm c}$ for this source is 7.3\,TeV. By adjusting the physical parameters to make $\tau_{\rm \gamma\gamma}(E_{\rm c})$ equal to 1, we have $f_{\rm ph}\simeq 10^{-3}$ at the proton energy $E_{p,c}\simeq 3\times 10^5E_{\rm c} \simeq 2.2\times 10^{18}$eV where $p-\gamma$ interaction efficiency reaches the maximum. Since $F(E_p)$ is determined by the photon spectra of low-energy SED, we can fully determine $f_{p\gamma} = f_{p\gamma}(E_{p,c}) F(E_p)$ for this source, as shown in Fig.~\ref{fig:efficiency}. One can in principle further increases the value of $f_{p\gamma}$ by adjusting certain parameters, but the gamma-ray opacity around $E_{\rm c}$ will then become larger than unity and we would expect a break or cutoff in the TeV spectra around $E_{\rm c}$, which however, is not seen in the data.


More quantitatively, for a BL Lac object with a hard TeV spectrum, we first correct for the influence of EBL attenuation on the spectrum. Then, we look for 
the highest energy data point after which a significant suppression or a softening appears in the spectrum, or simply the highest energy data point if no 
spectrum suppression or softening appears. The corresponding energy of the data point has been defined as $E_{\rm c}$. We then fit the low-energy bump in 
the SED of the BL Lac object with the synchrotron radiation of primary electrons and choose model parameters to achieve $\tau_{\rm \gamma\gamma}(E_{\rm c})=1$. 
Based on the resulting model parameters, we calculate the $p-\gamma$ interaction efficiency and denote the obtained value by $f_{p\gamma}^{\rm UL}$, 
where ``UL'' means that the obtained $p-\gamma$ efficiency is the upper limit for the source. The maximum beam-corrected luminosity 
(i.e., assuming that the inferred luminosity is emitted only in a beam of opening angle $\theta_j \sim 1/\Gamma$) of electromagnetic particles
produced in the $p-\gamma$ interactions can then be given by
\begin{equation}\label{eq:lph_ul}
L_{\rm ph}^{\rm UL} = \frac{4}{3}\pi R^3\delta_{\rm D}^2m_{\rm p}c^2\int \frac{5}{8}f_{\rm ph}^{\rm UL}(\gamma_{\rm p})\gamma_{\rm p}Q_{\rm p}(\gamma_{\rm p})d\gamma_{\rm p},
\end{equation}
for the photohadronic process where the factor $5/8$ considers about $3/8$ of the lost proton energy goes into neutrinos, and
\begin{equation}\label{eq:lbh_ul}
L_{\rm BH}^{\rm UL} = \frac{4}{3}\pi R^3\delta_{\rm D}^2m_{\rm p}c^2\int f_{\rm BH}^{\rm UL}(\gamma_{\rm p})\gamma_{\rm p}Q_{\rm p}(\gamma_{\rm p})d\gamma_{\rm p}.
\end{equation}
for the BH process. We obtain the total beam-corrected luminosity from the $p-\gamma$ processes as $L_{p\gamma}^{\rm UL}=L_{\rm ph}^{\rm UL}+L_{\rm BH}^{\rm UL}$.
Regardless of the details of the electromagnetic cascade induced by those electromagnetic particles, we have $L_{\rm TeV} < L_{p\gamma}^{\rm UL}$ simply from the 
perspective of the energy budget where $L_{\rm TeV}$ is the intrinsic (i.e., beam-corrected) TeV gamma-ray luminosity that cannot be explained by 
leptonic processes, since the TeV emission eventually originates from the electromagnetic particles generated in the $p-\gamma$ model as we mentioned earlier. On the other hand, according to the expression for the kinetic power in relativistic protons $P_{\rm p, k}$, i.e., Eq.~(\ref{eq:Ppk}), we can see that 
the ratio of $L_{p\gamma}$ to $P_{\rm p, k}$ does not depend on the proton luminosity, but on the interaction efficiency and the spectral shape of injected 
protons (i.e., power-law index $q$ and the cutoff energy). Thus, to account for the TeV emission through the $p-\gamma$ processes, the required nonthermal 
proton injection power of the jet should satisfy
\begin{equation}
P_{\rm p, k}\geq L_{\rm TeV}\left[\frac{4\int^{\gamma_{p,\rm max}}_1\left(\frac{5}{8}f_{\rm ph}+f_{\rm BH}\right)\gamma_p^{1-q}d\gamma_p}
{3\int^{\gamma_{p,\rm max}}_1\gamma_p^{1-q}d\gamma_p}\right]^{-1}
\end{equation}
since $L_{\rm TeV}\leq L_{p\gamma}^{\rm UL}$.
We note that $\tau_{\gamma\gamma}(E_{\rm c})$ is not necessarily equal to unity. The cutoff or softening feature in TeV spectrum may be simply due to the cutoff 
or softening in the spectrum of the emitting particles instead of $\gamma\gamma$ absorption. Therefore, the realistic value of $f_{p\gamma}$ can be even 
smaller than the one obtained by imposing $\tau_{\gamma\gamma}(E_{\rm c}) = 1$.

\begin{figure*}
\centering
\includegraphics[angle=0.2, scale=0.31]{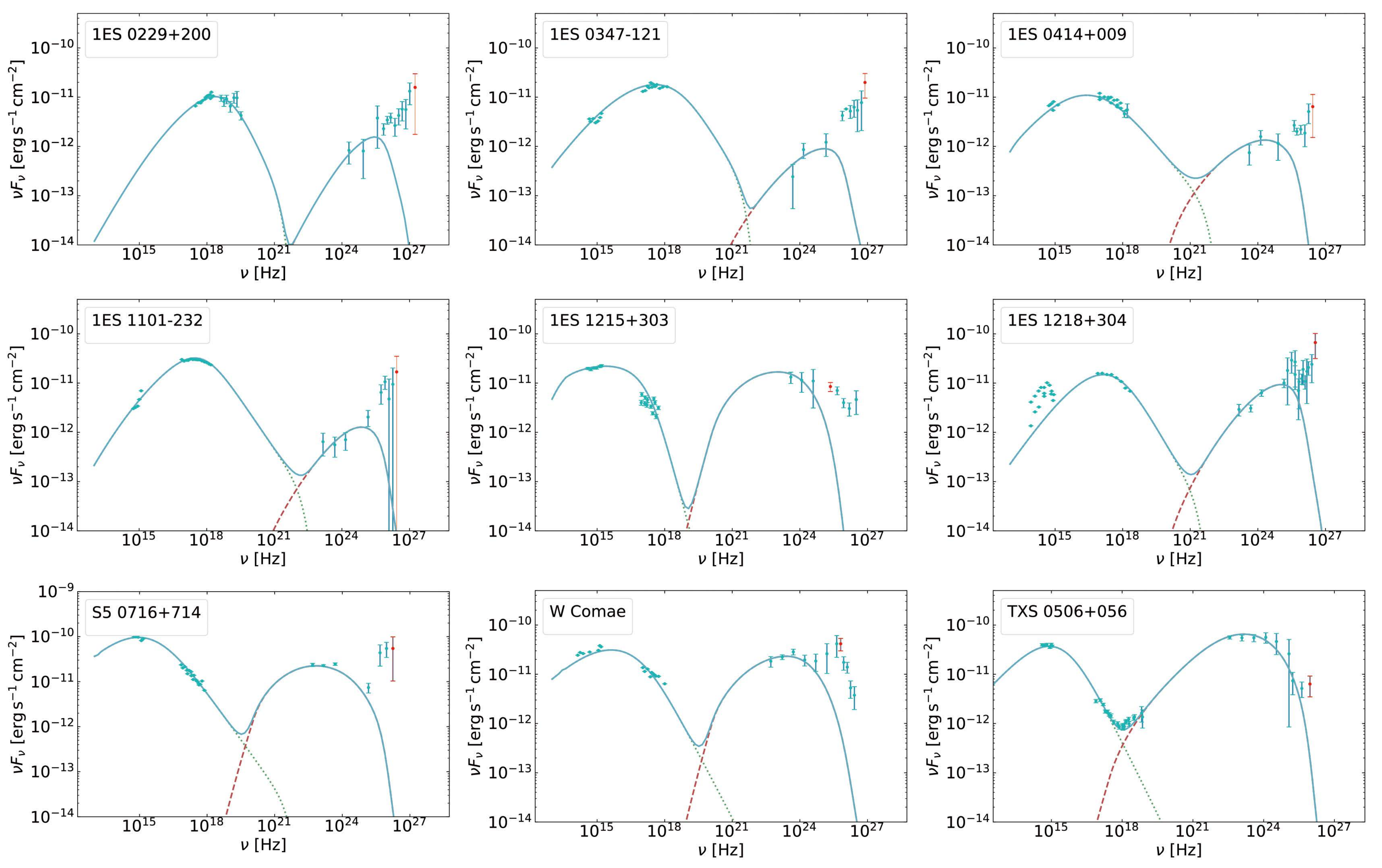}
\caption{The SSC modeling of the SEDs of TeV BL Lacs. The blue data points are quasi-simultaneous data from the literature. The internal 
$\gamma\gamma$ absorption optical depth $\tau_{\gamma\gamma}$ of the red data point is 1. The TeV data points show the intrinsic emission which 
are corrected with the EBL model specified in the respective references.}\label{fig:sed}
\end{figure*}

\section{Applications}\label{app}
We now apply the procedure introduced in the preceding section to some TeV BL Lac objects. To reduce the uncertainty caused by the model parameters 
as much as possible, we select our sample of BL Lac objects according to the following criteria: (i) the redshift is known; (ii) (quasi-)simultaneous 
multi-wavelength SED data are available; (iii) the TeV emission of the sources can not be well reproduced by the leptonic model or at least the origin 
of the TeV emission is under debate in the previous literature (see the Appendix.\ref{appendix:ana} for details). Note that SEDs of these BL Lac objects 
may still be fitted by the leptonic model (or other models different from $p-\gamma$ models) via introducing geometry effects or multiple emission 
zones, but we here consider the conventional hadronic scenario and examine the requirement on the jet power in the framework of $p-\gamma$ models.

Based on the above criteria, we collect a sample of 3 IBLs and 6 HBLs  (according to the classification by \cite{2010ApJ...716...30A}) from the 
TeVCat. The (quasi-)simultaneous SEDs of 1ES 0229+200, 1ES 0347-121, 1ES 0414+009, 1ES 1101-232, 1ES 1215+303, 1ES 1218+304, S5 0716+714, W Comae and TXS 0506+056 are taken from \cite{2014ApJ...782...13A}, \cite{2007A&A...473L..25A}, \cite{2012A&A...538A.103H}, \cite{2007A&A...470..475A}, 
\cite{2012A&A...544A.142A}, \cite{2010MNRAS.401..973R}, \cite{2009ApJ...704L.129A}, \cite{2009ApJ...707..612A} and \cite{2018arXiv180708816I}, respectively. It should be noted that there is no simultaneous GeV data for 1ES 0347-121, 1ES 1101-232, 1ES 1218+304, S5 0716+714 during the TeV observation since the TeV observation is performed before the launch of \textit{Fermi} satellite. We then use the two-year average {\it Fermi}-LAT data (2008.08.04 - 2010.08.01)\footnote{The database of the Space Science Data Center (SSDC) SED Builder provides the two-year average {\it Fermi}-LAT data; https://tools.ssdc.asi.it/SED/} in our fitting. In addition, the quasi-simultaneous GeV data of W Comae are taken from \cite{2010ApJ...716...30A}. In the SED of 1ES 1218+304, the thermal radiation is prominent in the infrared band which is believed to originate from the host galaxy \citep{2010MNRAS.401..973R}, and hence we do not consider the interaction processes (including IC radiation, $\gamma\gamma$ annihilation and $p-\gamma$ interactions) on this thermal component. Note that $f_{p\gamma}$ and $\tau_{\gamma\gamma}$ caused by this thermal component still follow the aforementioned relation, so we do not expect that including this external photon field can significantly increase $f_{p\gamma}$ even if this component is from a much more compact region with a larger photon density. We also note that the VHE emission of 1ES 1215+303, 1ES 1218+304, W Comae and TXS 0506+056 show significant variation, while no evidence of variability in TeV is found for the other sources. Our method is based on the relation between the $\gamma\gamma$ annihilation opacity and the $p-\gamma$ interaction efficiency. Thus, the method can be generalized to both high and low state of the blazar as long as these two processes operate in the same target photon field which is considered to be the synchrotron radiation of accelerated electrons.

\begin{figure*}[htbp]
\centering
\includegraphics[angle=0.2, scale=0.30]{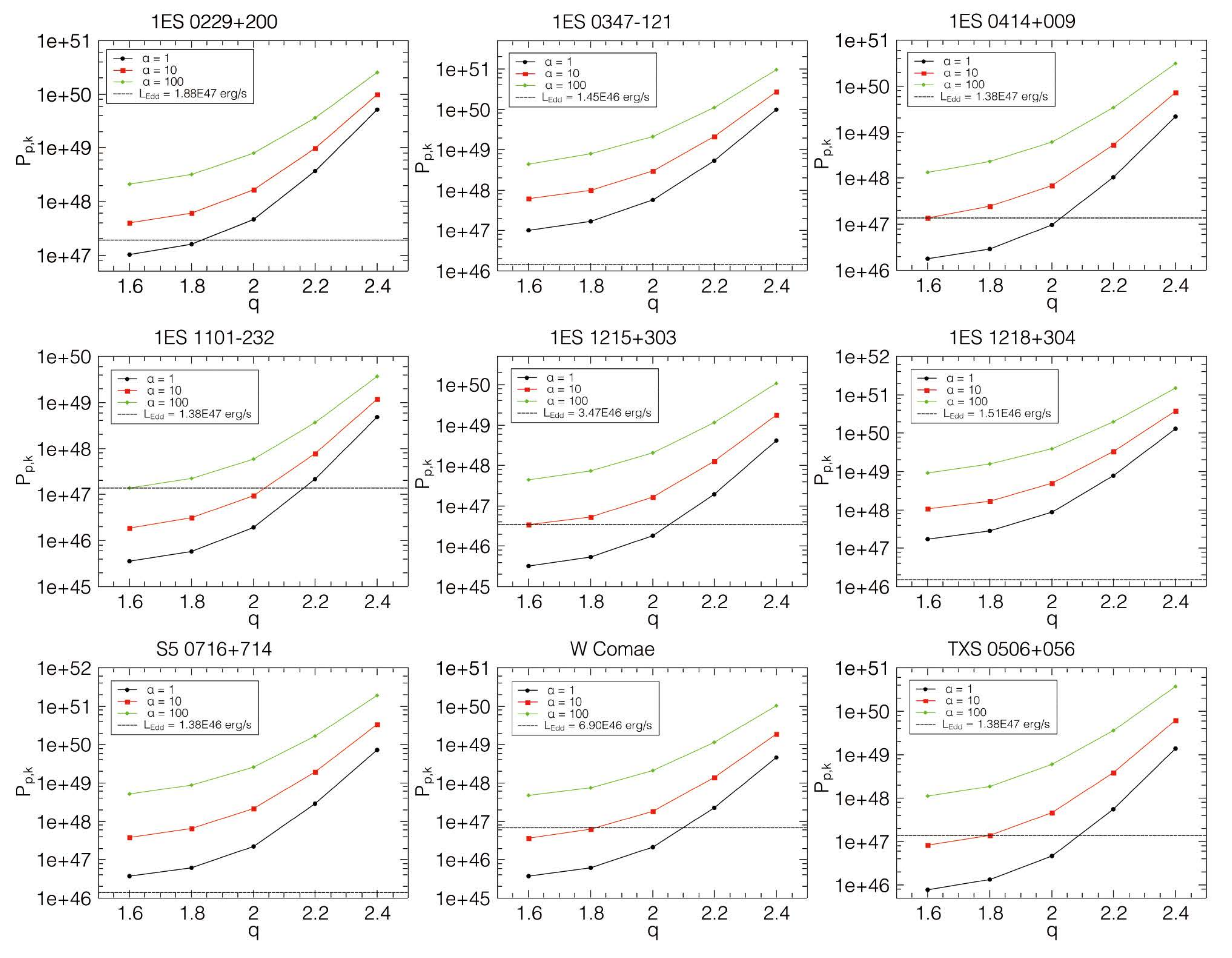}
\caption{The kinetic power in relativistic protons with different values of $\alpha$ and $q$. The black hole masses used to calculate the 
Eddington luminosity can be found in Table~\ref{tab:uhbl}. }\label{fig:Lmin}
\end{figure*}

Using the standard one-zone leptonic SSC model \citep{2001A&A...367..809K}, we firstly fit the low-energy bump with synchrotron radiation of primary electrons for each BL Lac in the sample. Instead of exploring the entire parameter space to optimize the fitting, we look for parameters to achieve a $\gamma\gamma$ annihilation opacity equal to unity at the critical energy (i.e., $\tau_{\gamma\gamma}(E_{\rm c}) = 1$ as shown with red data points in Fig.~\ref{fig:sed}), such that the upper limit of the $p-\gamma$ interaction efficiency can be obtained. On the premise of a reasonable fitting to the low-energy bump, we also try to fit the high-energy data with SSC emission as much as possible. The fitting results are shown in Fig.~\ref{fig:sed} and the model parameters are shown in Table~\ref{tab:para}. One can see that the SEDs of these 9 BL Lacs from optical to the GeV band can be fitted well by the leptonic model, but it starts to fail for photons above 100\,GeV. We then calculate the total luminosity beyond that energy and obtain $L_{\rm TeV}${\footnote{The integrated TeV luminosity is estimated by the following methods: Given a series of data points of the energy and the flux $\{(E_0,F_0),(E_1,F_1),\cdot \cdot \cdot,(E_i,E_i),\cdot\cdot\cdot ,(x_n,y_n)\}$ ($E_0<E_1<\cdot\cdot\cdot<E_n$) which are not fitted by the leptonic model. We calculate the beam-corrected luminosity in the range of $E_0-E_n$ by the trapezoidal rule, i.e., $ L_{\rm TeV} = 4\pi D_L^2\sum_{i=0}^{n-1} (F_i+F_{i+1})(E_{i+1}-E_i)/2\delta_D^2$ where $D_L$ is the luminosity distance of the source. Error bars are not considered.}.

Considering the obtained leptonic emission as the target photon field, we calculate the $p-\gamma$ interaction efficiency via Eq.~\ref{eq:eff_ph} 
and Eq.~\ref{eq:eff_bh}. To study the influence of the proton injection spectrum, we employ several values for the spectral index $q$ in the range 
of $1.6-2.4$ and three different $\gamma_{\rm{p, max}}$ with $\alpha=1$, $\alpha=10$ and $\alpha=100$ respectively. The resulting minimum $P_{\rm p, k}$ for different BL Lacs in each combination of $q$ and $\gamma_{p,\rm max}$ are shown in Fig.~\ref{fig:Lmin}. We can see that the minimum $P_{\rm p, k}$ decreases as $q$ and $\alpha$ become smaller. This is because given a harder injection spectrum (i.e., a smaller $q$) and a higher cutoff energy in the spectrum (i.e., a smaller $\alpha$), more energy is distributed to high energy where the $p-\gamma$ interaction efficiency is larger. For most sources, the minimum $P_{\rm p, k}$ is larger than the corresponding Eddington luminosity $L_{\rm Edd}$ for most combinations of $\alpha$ and $q$. Particularly, even with a very hard injection proton spectrum (i.e., $q=1.6$) and with the extreme case of $\alpha=1$, $P_{\rm p, k}$ is still larger than the Eddington luminosity for 1ES 0347-121, 1ES 1218+304, and S5 0716+714, suggesting that a super-Eddington jet luminosity is needed in the $p-\gamma$ model.

We note that the minimum jet powers obtained in this work are conservative. First, a considerable fraction of proton energy lost into EM particles in 
$p-\gamma$ interactions may be reprocessed into the X-ray or lower energy band via synchrotron radiation of the generated electrons/positrons so that we 
need a larger $L_{p,\rm inj}$ to account for the TeV emission. Second, we intentionally choose parameters to achieve a gamma-ray opacity equal to unity 
at the critical energy $E_{\rm c}$ when we fit the low-energy bump in the SED of BL Lacs in the sample. However, as mentioned in the preceding discussion, 
the opacity can be much smaller than unity and hence the $p-\gamma$ interaction efficiency used in this work is most likely an overestimation. In addition, 
we do not expect the jet to consist only of relativistic protons. Usually, one would expect the jet to contain more cold protons than relativistic protons 
and thus the obtained relativistic proton power may only constitute a small part of the jet power.

\begin{table*}
\caption{Model parameters for SED fitting. We set $\gamma_{\rm e, max} = 10^7$ for all sources which will not affect our fitting results. The first six 
objects are HBLs, the last three objects are IBLs.}
\centering
\begin{tabular}{@{}lrrrrrrrrl@{}}
\hline\hline
object	&	R &	B	&	$\delta_{\rm D}$	&	$q_1$	&	$q_2$	&	$\gamma_{\rm e, min}$	&	$\gamma_{\rm e, break} $ & $L_{e, \rm inj}$ & $E_{\rm c}$	\\
    &   (cm) & (Gauss) & & & & & & (erg/s) & (TeV)  \\
\hline
1ES 0229+200	&	3.98$\times10^{15}$   	&	0.43	&	12.71	&	1.32	&	3.95	&	  3.02$\times10^2$	&	  5.62$\times10^5$	& 1.55$\times10^{41}$ & 7.33\\
1ES 0347-121	&	 1.21$\times10^{16}$   	&	0.35	&	28.79	&	1.83	&	3.5	&	  3.14$\times10^2$	&	  2.82$\times10^5$	& 3.40$\times10^{40}$ & 3.21\\
1ES 0414+009	&	 7.08$\times10^{15}$   	&	0.75	&	29.13	&	1.82	&	3.1	&	  3.11$\times10^2$	&	  3.98$\times10^4$	& 5.64$\times10^{40}$ & 1.13\\
1ES 1101-232	&	 3.16$\times10^{15}$   	&	3.60	&	24.02	&	1.41	&	3.35	&	  1.00$\times10^1$	&	  4.75$\times10^4$	& 6.89$\times10^{40}$ & 1.11\\
1ES 1215+303	&	 1.58$\times10^{15}$   	&	3.10	&	15.04	&	2.24	&	5.45	&	  3.08$\times10^2$	&	  2.51$\times10^4$	& 2.73$\times10^{41}$ & 0.09 \\
1ES 1218+304	&	 5.98$\times10^{15}$   	&	0.36	&	19.81	&	1.68	&	3.6	&	  3.00$\times10^2$	&	  1.32$\times10^5$	& 1.33$\times10^{41}$ & 1.47\\
S5 0716+714	&	 2.59$\times10^{16}$   	&	0.53	&	27.97	&	2.02	&	3.4	&	  1.00$\times10^1$	&	  1.02$\times10^4$	& 1.61$\times10^{42}$& 0.73\\
W Comae	&	 2.00$\times10^{15}$   	&	2.13	&	16.36	&	2.11	&	3.75	&	  1.00$\times10^0$	&	  2.40$\times10^4$	& 2.97$\times10^{41}$ & 0.26\\
TXS 0506+056	&	 8.91$\times10^{16}$   	&	0.11	&	16.51	&	1.92	&	4.07	&	  1.00$\times10^1$	&	  1.91$\times10^4$	& 9.21$\times10^{42}$ & 0.38\\
\hline
\end{tabular}\label{tab:para}
\end{table*}

\begin{table*}
 \caption{Parameters relevant to the jet powers of selected BL Lacs. $z$ is the redshift of the source; $P_{\rm e, k}$ is the kinetic power in relativistic 
 electrons in the AGN frame in unit of erg/s; Log$M_{\rm BH}$ is the logarithm of the SMBH in units of solar masses, $M_\odot$; $L_{\rm TeV}$ 
 is the intrinsic beam-corrected luminosity of the TeV data in units of erg/s; $P_{\rm p, k}/L_{\rm Edd}$ is the ratio of the minimum injection proton 
 luminosity to the Eddington luminosity in the case of $\alpha$ = 10 and $q$ = 2; $P_{\rm p, k}/P_{\rm e, k}$ is ratio of the minimum injection proton 
 luminosity to the injection electron luminosity in the case of $\alpha$ = 10 and $q$ = 2. For 1ES 0414+009, 1ES 1101-232 and TXS 0506+056, in absence of 
 an estimated black hole mass, we considered an average value of $10^9 M_\odot$ \citep{2017ApJ...851...33P}.}
       $$
   \begin{array}{ p{65 pt} p{30 pt}  p{80 pt}  p{50 pt} p{45pt} p{55pt}p{55pt} }
            \hline
             \noalign{\smallskip}
              object	&	$z$	&	 Log$M_{\rm BH}$	&	$P_{\rm e, k}$ 	&	$L_{\rm TeV}$	&	$P_{\rm p, k}/L_{\rm Edd}$	&	$P_{\rm p, k}/P_{\rm e, k}$ 	\\
               & & ($M_\odot$)  & (erg/s)  & (erg/s) & ($\alpha = 10, q = 2$) & ($\alpha = 10, q = 2$) \\
            \noalign{\smallskip}
            \hline
            \noalign{\smallskip}
1ES 0229+200	&	0.14	&	9.16 $\pm$0.11 \newline \citep{2012AA...542A..59M}	&	2.51$\times10^{43}$	&	6.21$\times10^{42}$	&	8.73$\times10^0$	&	1.49$\times10^5$	\\
1ES 0347-121	&	0.188	&	8.02 $\pm$ 0.11 \newline \citep{2012AA...542A..59M}	&	2.82$\times10^{43}$	&	1.97$\times10^{42}$	&	2.05$\times10^2$	&	1.05$\times10^5$	\\
1ES 0414+009	&	0.287	&	9	&	4.79$\times10^{43}$	&	1.82$\times10^{42}$	&	5.04$\times10^0$	&	1.45$\times10^4$	\\
1ES 1101-232	&	0.186	&	9	&	3.98$\times10^{43}$	&	1.76$\times10^{42}$	&	6.84$\times10^{-1}$	&	2.37$\times10^3$	\\
1ES 1215+303	&	0.13	&	8.4 \newline \citep{2012NewA...17....8G}	&	6.17$\times10^{43}$	&	9.34$\times10^{42}$	&	4.79$\times10^0$	&	2.69$\times10^3$	\\
1ES 1218+304	&	0.182	&	8.04 $\pm$ 0.24 \newline \citep{2012AA...542A..59M}	&	5.21$\times10^{43}$	&	1.39$\times10^{43}$	&	3.29$\times10^2$	&	9.56$\times10^4$	\\
S5 0716+714	&	0.31	&	8 \newline \citep{2015MNRAS.450L..21Z}	&	1.26$\times10^{45}$	&	4.26$\times10^{43}$	&	1.58$\times10^2$	&	1.73$\times10^3$	\\
W Comae	&	0.101	&	8.7 \newline \citep{2015MNRAS.450L..21Z}	&	7.94$\times10^{43}$	&	2.24$\times10^{42}$	&	2.70$\times10^0$	&	2.34$\times10^3$	\\
TXS 0506+056	&	0.3365	&	9	&	2.51$\times10^{45}$	&	6.40$\times10^{42}$	&	3.33$\times10^0$	&	1.83$\times10^2$	\\
           \noalign{\smallskip}
            \hline
         \end{array}
      $$
        \label{tab:uhbl}
  \end{table*}

\section{Discussion and conclusions}\label{dis}

The Eddington luminosity is obtained by balancing the force of radiation pressure and gravity of an object. Although it is not a strict limit on the luminosity of a black hole, the Eddington luminosity is usually regarded as a reasonable approximation for the maximum jet power of a blazar. Among the radiation models for blazar jets, the leptonic models usually require a sub-Eddington jet power since the radiation efficiency of electrons is high. The low radiation efficiency of protons in the hadronic models (either the $p-\gamma$ model or the proton synchrotron model) obtained in this work and previous studies \cite{2011IAUS..275...59S, 2015MNRAS.450L..21Z}, implies a super-Eddington jet power. 
Such a jet may be powered by other mechanism such as the Blandford-Znajek mechanism \citep{1977MNRAS.179..433B} which extracts the spin power of the SMBH or by the super-critical accretion. In the former scenario, however, \citet{2015MNRAS.450L..21Z} pointed out that the magnetic fluxes measured through the radio-core shift effect in some blazars rule out the later mechanism. The latter scenario, i.e., super-critical accretion onto SMBHs, has been studied in various works \citep[e.g.][]{1998MNRAS.297..739B, 2015ApJ...804..148V, 2015MNRAS.453.3213S}. \citet{2015MNRAS.453.3213S} found that powerful jets with super-Eddington luminosity may be able to launch from the SMBH only under some uncommon conditions (such as in a tidal disruption event). However, even if the condition can be satisfied, the super-critical accretion mode can only last a very small fraction of the lifetime of a SMBH as indicated by \cite{2015MNRAS.450L..21Z}, otherwise the growth of the SMBH would be too quick. Thus, such an accretion mode can be only applied to a tiny fraction of blazars. Furthermore, simulation \citep{2016MNRAS.456.3929S} shows that the radiation in the funnel along the axis is supposed to be super-Eddington (which is the case of BL Lacs) when the accretion is super-critical. However, from the non-detection of the spectral feature of the accretion-disk emission in the SED of the BL Lacs in our sample, we can estimate upper limits of the disk luminosity for these sources to be $10^{44}-10^{45}$erg/s which are sub-Eddington. On the other hand, in the picture of jet/disk symbiosis \citep{Falcke95}, although it is possible that the accretion power are channelled into the kinetic energy of the jet or the wind than into the disk radiation, the theoretical expectation for the ratio between the jet's kinetic luminosity and the accretion-disk luminosity is $\lesssim 10$ for a large range of reasonable parameters \citep{Donea96, Donea03}. It is much smaller than the ratio required in the $p-\gamma$ model for BL Lacs in our sample which is $>10^3-10^4$. Thus, null detection of the accretion-disk emission from these BL Lacs disfavor super-critical accretion in these sources. Besides, jet powers estimated from radio lobes and X-ray cavities \citep{2007MNRAS.381..589M, 2012Sci...338.1445N} are in conflict with the required super-Eddington jet power at a timescale of $1-10\,$Myr. 
If the $p-\gamma$ model applies, it probably implies a different picture for the accretion of SMBH in blazars than the one depicted by the standard theory.

\citet{2016ApJ...831..142M} suggest that the jet power can be reduced significantly by introducing a huge amount of positrons to replace protons (in their case, the number density of positrons is 30 times higher than that of protons) in the jet from the point of view of keeping the neutrality of the jet. We note that although this is a possible solution to some sources, it does not apply to the BL Lacs in our sample. This is because the IC radiation of positrons also suffers the KN suppression and hence cannot explain the hard TeV spectrum (at least in the one-zone model).

To summarize, we obtained a conservative yet robust lower limit on the jet power for TeV BL Lacs for which the standard leptonic model does not work well. The detection of TeV photons from BL Lacs imposed an upper limit for the $\gamma\gamma$ annihilation opacity. Since $p-\gamma$ interactions (including photopion production and the BH process) take place in the same target photon field as $\gamma\gamma$ annihilation, the $p-\gamma$ interaction efficiency is linked with the $\gamma\gamma$ opacity. Based on this relation, we obtained an upper limit for the $p-\gamma$ interaction efficiency which translates to the minimum proton power of the jet if $p-\gamma$ interactions are responsible for the TeV emission from these BL Lacs. By applying this approach to a sample of 9 TeV BL Lacs, we found that the minimum injection proton power is larger than the Eddington luminosity for most combinations of $q$ and $\alpha$. If the Eddington luminosity is the maximum luminosity that a blazar jet can achieve, the $p-\gamma$ process may not be responsible 
for the TeV emission in these TeV BL Lacs. One then may have to consider the leptonic origin with more complicated topology of the radiation zone. On the other hand, the radiation efficiency of protons in the hadronic model that employs the hadronuclear interaction is not related to the gamma-ray opacity \citep[e.g.][]{1997ApJ...478L...5D, 1997MNRAS.287L...9B, 2010A&A...522A..97A, 2010ApJ...724.1517B, 2013ApJ...774..113K, 2018arXiv180705113L, 2018arXiv180805651S}, and hence may provide a solution to fit the TeV spectrum with a sub-Eddington jet power.


%

\acknowledgments
Acknowledgments.

\appendix
\section{Difficulties of the standard leptonic models in fitting TeV BL Lacs}\label{appendix:ana}
We here summarize the difficulties of the leptonic interpretation of the TeV emissions from the sources in our selected sample, based on results and 
arguments in previous literature:

(1) A large value of the Doppler factor ($\delta_{\rm D}$) is needed. For 1ES 0229+200, one of the important outcome of the SSC interpretation 
\citep{2014ApJ...782...13A} is that the minimum $\delta_{\rm D}$ required in the fitting of SSC model is $\delta_{\rm D} \geqslant 53$, which is 
significantly higher than the commonly adopted value for blazar jets. In earlier studies, a large $\delta_{\rm D}$ has been commonly suggested for 
this object, e.g., \cite{2009MNRAS.399L..59T} adopt $\delta_{\rm D}$ = 50 and \cite{2011A&A...534A.130K} adopt $\delta_{\rm D}$ = 40. Similarly, 
for 1ES 0414+009, \cite{2012A&A...538A.103H} adopt $\delta_{\rm D} = 40$ with either a SSC model and an EC model. Even with such a large 
$\delta_{\rm D}$, the SED fitting is only marginal. For 1ES 1215+303, $\delta_{\rm D} = 60$ is required \citep{2012A&A...544A.142A}. For 
1ES 1218+304, an extreme value of $\delta_{\rm D}$ = 80 is employed \citep{2010MNRAS.401..973R}. An extremely high $\delta_{\rm D}$ imply 
a very fast movement of the radiation region, though such a high $\delta_{\rm D}$ is inconsistent with radio observations of the movement 
of knots \citep{2013AJ....146..120L} or with the statistics of blazars \citep{2006ApJ...640..185H}.

(2) A high value of $\gamma_{\rm e, min}$ is needed ($\sim 10^4$) in the SED fittings for 1ES 0229+200, 1ES 0347-121, 0414+009, 1ES 1101-232 
and 1ES 1215+303. It requires some specific conditions in those sources to make such fine-tuned values of $\gamma_{\rm e,min}$ physically
reasonable \citep{2006MNRAS.368L..52K}.

(3) The {\it Fermi}-LAT observation is hard to reconcile with the VHE observation under the leptonic model for some sources. For S5 0716+714, 
either fitting with the one-zone leptonic model or with the spine-sheath model, the predicted gamma-ray flux overshoot the observed flux by 
{\it Fermi}-LAT. Moreover, an extremely high $\gamma_{\rm e, min} = 10^4$ is also needed \citep{2009ApJ...704L.129A}. For W Comae, SED was 
initially well fitted with both the SSC model and the EC model by \cite{2009ApJ...707..612A} without the {\it Fermi} observation. However, 
the leptonic model fails to fit SED after the GeV data \citep{2013ApJ...768...54B} is included.

(4) If a high-energy neutrino event is coincident both temporally and spatially with a $\gamma$-ray flare from a blazar such as in the case 
of TXS 0506+056 \citep{2018Sci...361..147I}, hadronic processes has to be considered. Therefore, TXS 0506+056 is included in our sample.

We here take 1ES 0229+200 as an example to show why the leptonic model can not fit the hard TeV spectrum. The synchrotron peak frequency of 
electrons can be estimated as $\nu_{\rm syn} \approx 3\times 10^6 \gamma_{\rm e, b}^2 B \delta_{\rm D}/(1 + z)$. Meanwhile, the SSC spectrum peaks at $\nu_{\rm ssc} \approx 4\gamma_{\rm e, b}^2 \nu_{\rm syn}/3$ which is emitted by electrons of $\gamma_{\rm e,b}$ scattering off the photons from the peak of the synchrotron bump, if we do not consider the KN effect. 
 Thus, in principle one can ascribe the hard TeV spectrum to the SSC emission as long as $h\nu_{\rm ssc}\geq E_{\rm c}$ which is generally true for IBL and HBL. But one also need to guarantee that KN effect does not interfere the spectrum below $E_{\rm c}$, i.e., $\nu_{\rm KN}\simeq \delta_{\rm D}\gamma_{\rm e,b}g(\alpha_1, \alpha_2)m_{\rm e}c^2/h(1+z)>E_{\rm c}$ where $g(\alpha_1,\alpha_2)={\rm exp}\left[1/(\alpha_1-1)+1/2(\alpha_2-\alpha_1))\right]\lesssim 1$ \citep{1998ApJ...509..608T} with $\alpha_1$ and $\alpha_2$ the spectrum index (i.e., $f_\nu\propto \nu^{-\alpha_{1,2}}$) of the synchrotron emission below and above the peak respectively. It translates to
\begin{equation}
\delta_{\rm D}>12g^{-2}(\alpha_1, \alpha_2)\left(\frac{B}{1\,\rm G}\right)\left(\frac{\nu_{\rm syn}}{10^{18}\rm Hz}\right)^{-1}\left(\frac{E_{\rm c}}{1\,\rm TeV} \right)^2
\end{equation}
Take 1ES 0229+200 for example, the observational synchrotron peak frequency is $\nu_{\rm syn} \approx 10^{18}$Hz, $E_{\rm c}\simeq 7\,$TeV and $g(\alpha_1,\alpha_2)\simeq 0.3$. Thus, to fulfill the above relation, we need $\delta_{\rm D}>5000(B/1\rm\, G)$. If we consider the typical magnetic field $B=0.1-1\,$G, the required $\delta_{\rm D}$ is $>100-1000$ which far exceeds the typical value. 
If we want to fit the TeV flux with the SSC radiation, we need to impose $\nu_{\rm KN}$ to the TeV range. A large $\gamma_{\rm e, b}$ 
has to be assumed, however, at the expanse of a poor fitting to the low-energy bump (as shown in Fig.~\ref{fig:KN}).

\begin{figure}
\centering
\includegraphics[angle=0.2, scale=0.8]{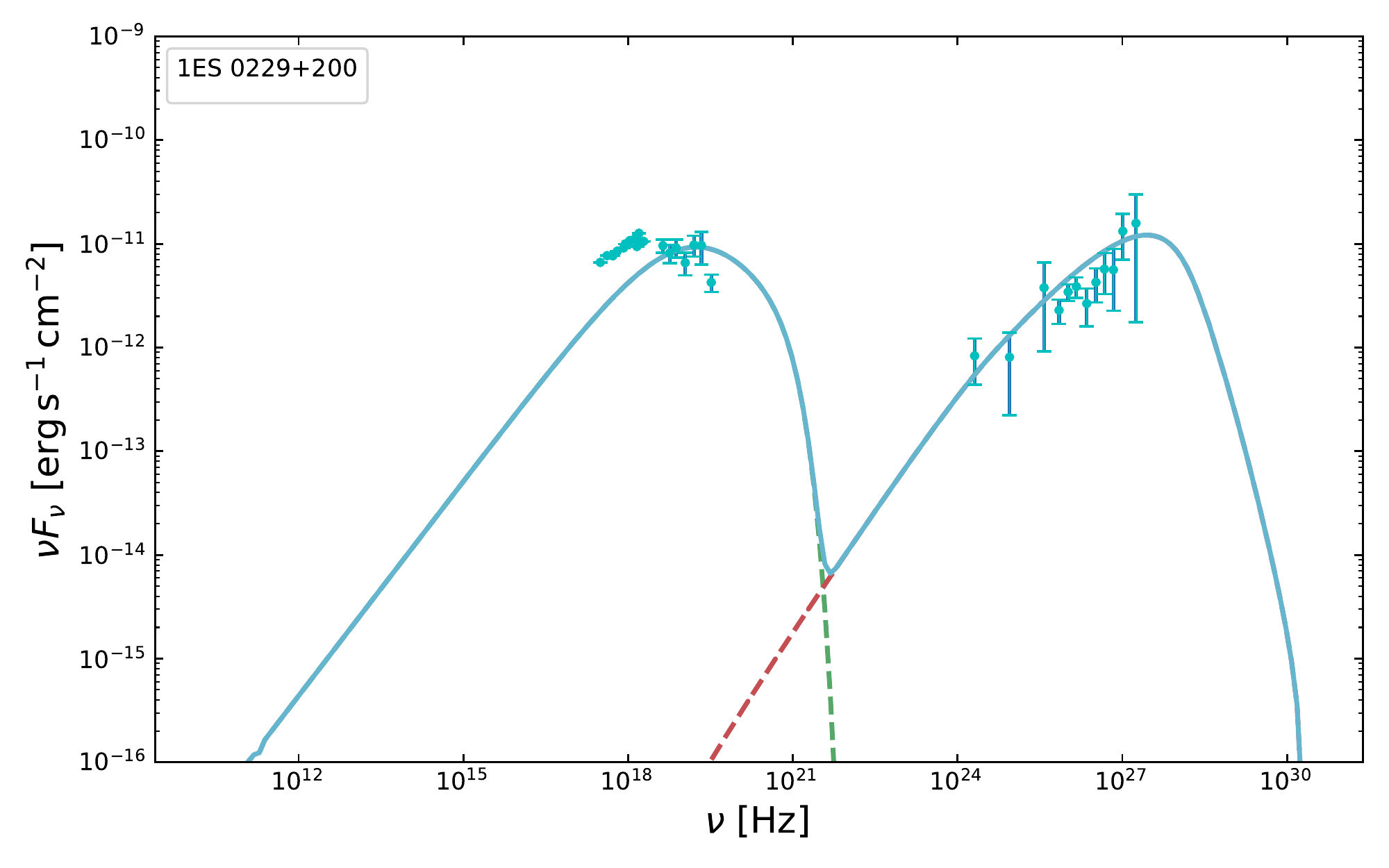}
\caption{The difficult in fitting the SED of TeV BL Lacs with the standard leptonic model. If the TeV spectrum is fitted, the low energy bump cannot be fitted. Model parameters: R = 1.18$\times10^{16}$ cm, B = 0.003 G, $\delta_{\rm D}$ = 18.71, $q_{\rm 1}$ = 1.62, $q_{\rm 2}$ = 3.95, $\gamma_{\rm e, min}$ = 1, $\gamma_{\rm e, b} = 8.62\times10^6$, $\gamma_{\rm e, max} = 5\times10^7$, $L_{e, \rm inj} = 1.25\times10^{42}$erg/s. }\label{fig:KN}
\end{figure}

\end{document}